\documentclass[10pt,journal,compsoc]{IEEEtran}
\usepackage{amsmath,amsfonts}
\usepackage{algorithmic}
\usepackage{algorithm}
\usepackage{array}
\usepackage[caption=false,font=normalsize,labelfont=sf,textfont=sf]{subfig}
\usepackage{textcomp}
\usepackage{stfloats}
\usepackage{url}
\usepackage{verbatim}
\usepackage{graphicx}
\usepackage{cite}
\usepackage{soul}

\usepackage{tabu}                      
\usepackage{booktabs}                  
\usepackage[dvipsnames]{xcolor}
\usepackage[OT1]{fontenc} 
\usepackage{microtype}                 
\PassOptionsToPackage{warn}{textcomp}  
\usepackage{textcomp}                  
\usepackage{mathptmx}                  
\usepackage{times}                     
\usepackage{mathptmx}                  
\usepackage{listings}
\usepackage{enumitem}
\usepackage{tabularx}
\usepackage{makecell}
\usepackage{hyperref}
\usepackage{orcidlink}

\newcommand{\inlinehdr}[1]{\vspace{0.5ex}\noindent{\textbf{#1}}}

\newcommand{\revision}[1]{\textcolor{teal}{#1}}
\renewcommand{\revision}[1]{#1}

\begin{document}
%
\title{Design Concerns for Integrated Scripting and Interactive Visualization in Notebook Environments}
%
%
%

\author{Connor Scully-Allison, Ian Lumsden, Katy Williams, Jesse Bartels, Michela Taufer, Stephanie Brink, Abhinav Bhatele, Olga Pearce, and Katherine E. Isaacs
\thanks{C. Scully-Allison and K. E. Isaacs are with the SCI Institute, e-mail: cscullyallison@sci.utah.edu.}
\thanks{I. Lumsden and M. Taufer are with University Tennessee, Knoxville.}
\thanks{K. Williams is with Davidson College.}
\thanks{J. Bartels is with Rincon Research.}
\thanks{S. Brink and O. Pearce are with Larence Livermore National Laboratory.}
\thanks{A Bhatele is with University of Maryland, College Park.}
}

\markboth{Journal of \LaTeX\ Class Files,~Vol.~14, No.~8, August~2015}%
{Scully-Allison \MakeLowercase{\textit{et al.}}: Visualization+Scripting Design}

\maketitle

\begin{abstract}
Interactive visualization can support fluid exploration but is often limited to predetermined tasks. Scripting can support a vast range of queries but may be more cumbersome for free-form exploration. Embedding interactive visualization in scripting environments, such as computational notebooks, provides an opportunity to leverage the strengths of both direct manipulation and scripting. We investigate interactive visualization design methodology, choices, and strategies under this paradigm through a design study of calling context trees used in performance analysis, \revision{a field which exemplifies typical exploratory data analysis workflows with big data and hard to define problems.} 
We first produce a formal task analysis assigning tasks to graphical or scripting contexts based on their specificity, frequency, and suitability.
We then design a notebook-embedded interactive visualization and validate it with intended users. In a follow-up study, we present participants with multiple graphical and scripting interaction modes to elicit feedback about notebook-embedded visualization design, finding consensus in support of the interaction model. We report and reflect on observations regarding the process and design implications for combining visualization and scripting in notebooks.
\end{abstract}


\IEEEpeerreviewmaketitle

\section{Introduction}

\IEEEPARstart{I}{nteractive}
visualizations offer powerful support for discovering novel insights in large and complex data. However, these same visualizations primarily use direct manipulation, limiting the queries and data manipulations to what the visualization developers foresaw. Scripting offers enormous flexibility in specifying queries, but its output often lacks context or is unintuitive. Seeking to understand and leverage the strengths of both interaction modes to support complex exploratory data analysis, we conduct a design study in the domain of parallel performance analysis that carefully considers this \revision{trade-off} throughout the entire process, from design through evaluation. 

The driving visualization problem in our design study is supporting the exploratory analysis of program performance in large-scale parallel computing. In this domain, analysts have a high level of programming acumen and thus may benefit from solutions that support scripting and visualization. Additionally, our collaborators support a domain-specific data science library for this analysis and deploy new functionality to users via Jupyter notebook examples. Thus the setting and available users make the project conducive to studying visualization in a notebook-embedded space. \revision{Furthermore, parallel performance analysis exemplifies workflows and problems in other data science domains with the presence of large, high-dimensional datasets, large search spaces for optimization, and complex mappings of how changes in inputs affect measured outputs.}

\revision{Using guidance from the visualization community \cite{sedlmair2012design},} we investigate designing a visualization-scripting workflow for analyzing calling context trees (CCTs), a common type of data collected for performance analysis. We conduct interviews with front-line performance analysts and gather task-related data through weekly meetings with collaborators. 

In formulating our task analysis we characterize tasks in terms of their \revision{specificity and frequency, with the rationale that highly specific tasks are aided by the expressivity of scripting while frequent tasks can be aided by the fluidity of interactive visualization.} We \revision{thus} use this characterization to assign the tasks to scripting and visualization aspects of our design.
Using this augmented task analysis, we design a notebook-embedded, interactive CCT visualization to fit in a hybrid scripting workflow rather than standing alone. While visualizations are typically an endpoint in notebook-based analysis, our CCT visualization supports the transfer of data and state between Python scripting cells and the interactive visualization context. This means that the data in the notebook variables can reflect changes made to the data in the visualization and vice-versa. 

We validate our design with seven performance analysis experts. The participants completed our evaluation tasks successfully and derived additional insights about program performance. Using their initial feedback regarding the visualization-scripting workflow, we design a follow-up study to further explore interaction styles and possible pitfalls of the paradigm. We heard consensus among participants regarding their support for the combination of visualizing and scripting with automatic updating of the cells. However, we also observed difficulties in state management, suggesting further design is needed to make notebook state more explicit. 

Based on participant feedback and reflections on the design process, we find that our augmented task analysis, focusing on specificity and frequency of tasks, anticipated user needs for flexibility and workflow and provides an avenue for evolution in participant analysis workflows. We recommend visualization+scripting be considered in exploratory workflows with technical users and provide guidance for how to design visualizations with this dual modality in mind.

In summary, the primary contributions of this work are:
\vspace{-0.5ex}
\begin{itemize}
    \itemsep=0ex
     
    \item A generalizable process for developing a scripting-aware task analysis (Section~\ref{sec:taskapproach}) and demonstration applied in a specific tree visualization context (Section~\ref{sec:justtasks}),
    \item A design for a notebook-embedded calling context tree visualization (Section~\ref{sec:cctvis}), 
    \item \revision{Results of a} study (Section~\ref{sec:study}) on interaction modes and workflow options in a scripting-embedded visualization, and
    \item Guidance and reflections (Section~\ref{sec:reflections}) on the  intentional design of notebook-embedded hybrid visualization-scripting workflows. 

\end{itemize}

\section{Related Work}
\label{sec:related}

Computational notebooks such as Jupyter~\cite{jupyter}, R-Markdown~\cite{rmarkdown-manual}, and Observable~\cite{Observable} provide a literate~\cite{Knuth1984} environment combining executable code, its output, and narrative text and documentation. 

Recent works have explored the ability of notebooks to aid in general exploratory analysis workflows~\cite{chattopadhyay2020s, shrestha2021unravel, franccoise2021marcelle, vaithilingam2019bespoke}. Some focus on code primarily while others consider visualization but do not emphasize its potential as a more active component of a iterative analysis process.  
A recent survey~\cite{ono2021interactive} explains Jupyter-provided functionality for embedding HTML charts in computational notebooks and their potential for integrating bespoke visualizations into analytical workflows. 

In Jupyter, code is input into individually executable {\em cells}, which when run may produce output, including visualizations. While these visualizations may be used in analysis, there is often a manual step in which users must translate their findings back to code due to the separation between the visualization and code cells. Several works consider the design and implementation of paradigms for tighter integration of visualization in notebooks~\cite{b2, mage, carbrera, NotebookJS, wang2022nova, epperson2023dead}. Of them, B2~\cite{b2}, \texttt{mage}~\cite{mage}, and a Svelte extension~\cite{carbrera} support creating interfaces to transfer from the visualization back to scripting cells.

In designing B2~\cite{b2}, Wu et al. also observe that ``neither direct manipulation or coding is best suited to all tasks.'' \revision{They identify three} ``frictional'' barriers to designing integrated scripting and visualization systems\revision{referred to as semantic, temporal and layout gaps}. \revision{These three gaps describe technical issues hindering integration. The semantic gap describes difficulties in transferring state information between visualizations and code. The temporal gap describes the mismatch between the transience of visualizations and the persistence of code. The layout gap describes tensions between the linear structure of a notebook and common layouts of multi-view visualizations. The B2 library provides tools for eliminating these barriers and enabling successful implementation of hybrid visualizations in Jupyter notebooks.}  Our work complements theirs by proposing and demonstrating a method for determining \textit{which} tasks are suited to visualization or scripting, advancing the observations that not all tasks are \revision{equally suited to either modality.} We also present findings regarding applying visualization and scripting hand-off design choices proffered by Wu et al. with participants performing their typical, specific analysis tasks.

\revision{AutoProfiler~\cite{epperson2023dead} is a Jupyter extension that adds a floating view to the notebook layout to provide visual summaries of pandas dataframes. The authors present experiments with automatic or manual updating of the view, a facet of the temporal gap. We also examine this design choice, but in the context of a custom design for a specific workflow rather than a general heads-up display.} 

\revision{\texttt{mage}~\cite{mage} proposes a technical solution for passing state between scripting and direct manipulation cells. Kery et al. do not say which kind of a cell a task should be assigned to, but that ``The user is free to write code for some tasks or use direct manipulation to generate code for other tasks, as is most comfortable to them." While they do no explicitly explore the space of task assignment, they solicit user feedback on several discrete single-task visualizations they design, observing ``programming experts tended to reject some tool ideas that they felt they could do much faster by writing code." This proclivity motivates our work as many notebook users have strong programming skills and desire expressiveness when they already know what they want to do. As we designed for such an audience, we seek to formalize our understanding of how to assign tasks to scripting or visualization.}

Several designs in the explainable artificial intelligence (XAI) domain embed visualizations in notebooks, but the transfer of data and state is one-way: from scripting cells to visualizations only~\cite{xenopoulos2022calibrate, guedj2018pycobra, clarke2021appyters}. 
Emblaze~\cite{sivaraman2022emblaze} and CausalVis~\cite{guo2023causalvis} offer two-way passing as we discuss in this work. Emblaze's description and evaluation focused primarily was more on the visual design. CausalVis presents the design of both visualization and scripting primitives with design emphasis on fitting with existing tools in causal interface. Our process directly interrogates the scripting and visualization choices, contributing to further understanding for visual designs in this growing paradigm.

Beyond notebooks, other works~\cite{hartmann2008design,bigelow2016iterating,hempel2019sketch} have supported transfer between scripting and GUIs contexts. Hanpuku~\cite{bigelow2016iterating} enables an iterative edit loop between D3~\cite{bostock-d3-2011} and Adobe Illustrator. \textsc{Sketch-N-Sketch}~\cite{hempel2019sketch} supports SVG authoring through both direct manipulation and code. In contrast, we focus on the {\em design} of a visualization to fit within a scripting workflow.

For networks, libraries such as GUESS~\cite{adar2006guess}, JUNG~\cite{omadadhain2005jung}, and GraphViz~\cite{ellson2001graphviz} have supported a more scripting-centered approach to analysis and visualization. These are general approaches offering high flexibility but placing a higher expectation of visual design on the analyst to use directly. Our focus is on design choices assuming a mix of scripting and pre-designed interactive visualization.

\section{Domain Background: Profiling}
\label{sec:background}

{\em Performance} describes how well an application executes on computing resources, typically in terms of metrics such as time-to-completion or efficiency of resource use. For applications run on High Performance Computing (HPC) resources, poor performance limits the problem size or level of detail that can be computed. Critical applications like hurricane, medicine, energy simulations use these already oversubscribed HPC systems. Even small improvements in performance can free significant resources, enabling more science.

Performance analysis is a key step in optimizing an application. Developers collect and analyze performance data. The most simple performance metric is the total execution time. More detailed metrics can be used to capture performance of specific regions in the application code (e.g., functions). 

Performance analysis is an intrinsically exploratory activity that shares many characteristics with other types of exploratory data science workflows. The concept of ``poor'' performance is ill-defined and the space of potential performance problems is vast. Often, a performance problem can be hard to articulate or identify without iterating through a loop of collecting, transforming, and visualizing performance data. 

\textbf{Profiles and calling context trees.} A common form of performance data is a {\em profile}. A profile accumulates metrics associated with predetermined code regions. For example, a simple profile collects total time spent in each function. Our domain experts examined more detailed profiles that accumulate multiple metrics at each {\em calling context}. A calling context (or ``call path'') is the chain of calls leading to a particular function call.

Calling contexts can be structured into a {\em calling context tree} (CCT), an unordered prefix tree of all calling contexts. Each node represents an invocation of a particular code region with a unique calling context. Each calling context is represented as a path from the root of the tree to a given node. 

\textbf{Calling context visualization.} CCT visualization is typically done through bespoke graphical user tools or ad hoc using libraries like GraphViz~\cite{ellson2001graphviz}. Common idioms are node-link diagrams~\cite{weidendorfer2004tool, DeRose2007, Ahn2009STAT, Lin2010}, indented trees~\cite{geimer2010further, adhianto2010hpctoolkit, bell2003paraprof}, or icicle plots (``flame graphs'') and sunbursts~\cite{Adamoli2010Trevis, moret2010exploring, 7081872, Faust2022Anteater}, all with color encoding a single attribute.

Multiple attributes have been shown with indented tree+table idioms~\cite{adhianto2010hpctoolkit} and node-embedded charts~\cite{DeRose2007}. Icicle plots and sunbursts permit length and arc length to encode attributes. However, these encodings suggest child attributes sum to their parents, which is not true for all attributes our domain experts wish to examine. \revision{For example, `exclusive time', an attribute measuring the time taken to execute code within the function without external (child) function calls, frequently has children with values greater than their parent nodes.}

  Some performance analysis tasks require analyzing collections of CCTs~\cite{bergel2017visual}, e.g.,per-CPU CCTs. Visualizations of these collections have focused on showing distributions over aggregated trees~\cite{nguyen2016vipact, xie2018ccts, nguyen2019visualizing, kesavan2021scalable}. Comparison across per-CPU trees was a priority of our target users during interviews, and thus we focus on tasks exploring CCTs which have already been aggregated across CPUs.

{\em Hatchet}~\cite{bhatele2019hatchet} is an open-source library for CCT analysis. It provides a canonical pandas-based~\cite{reback2020pandas} data model for data produced by HPC profilers, allowing users to leverage modern data analysis ecosystems. 

Hatchet provides numerous operators for combining multiple trees and deriving per-node metrics such as {\em speedup} (the ratio of timing data between two CCTs) and {\em imbalance} (the variance across trees). These operations result in a single tree regardless of number of inputs. Thus the two-run comparisons we present here are representative of more numerous comparisons using the library. Hatchet also features a call path query language~\cite{Lumsden2022} for complex filters of CCT data, which we use in our workflow.

\section{Design Process}
\label{sec:designprocess}

We followed the iterative design study methodology of Sedlmair et al.~\cite{sedlmair2012design}, informed by the criteria for rigor proposed by Meyer and Dykes~\cite{meyer2019criteria}. We describe our process in these frameworks and our resulting data and task analyses. See the supplemental materials for supporting artifacts.

Our core research team includes two visualization experts and a Hatchet project manager who met weekly. We also met at least monthly with the larger Hatchet development team. The visualization team interviewed two regular Hatchet users, identified as {\em frontline analysts} during the design phase. The Hatchet project manager, who met more frequently with the frontline analysts, communicated needs and requests from them and also gathered example use cases and notebooks. 

\subsection{Data}
\label{sec:data}

Our data is a multi-variate tree network structure, with a fixed but arbitrary number of attributes greater than or equal to two. Tree sizes vary depending on the program being analyzed, the amount of parallelization, and the collection tool used. For example, our evaluation datasets are approximately 1500 nodes and 2700 nodes, respectively. For massively parallel runs these can scale up to tens of thousands of nodes, with tens of levels.

Though many tools collect CCTs per parallel thread, our interviews with frontline analysts did not discuss viewing the distribution across processes. Rather, they did their analyses with one tree that aggregates data across all processes. We thus focus on that aggregated data. 

\subsection{Visualization+Scripting Task Analysis Approach}
\label{sec:taskapproach}

From project inception the target delivery mechanism has been Jupyter notebooks. Development teams at our collaborators' organization are made aware of new features through sample notebooks available through a common web portal. Given this notebook environment, when analyzing tasks, we determined whether the visualization or scripting capabilities should support them. In making this determination, we considered three major aspects of the task: 

\textbf{Specificity} describes the complexity and/or concreteness of a task. Very complex tasks, like transformations with several steps or terms, are more specific as are tasks that require precise values. Implementing these in visualization would require complex interactions that may be more suited to the expressivity of scripting.

\textbf{Frequency} describes how often a task is expected to be performed. The less frequent a task is, the more it may be a special case, better handled by the flexibility of scripting. 

\begin{figure}[h]
    \centering
    \includegraphics[width=.95\columnwidth]{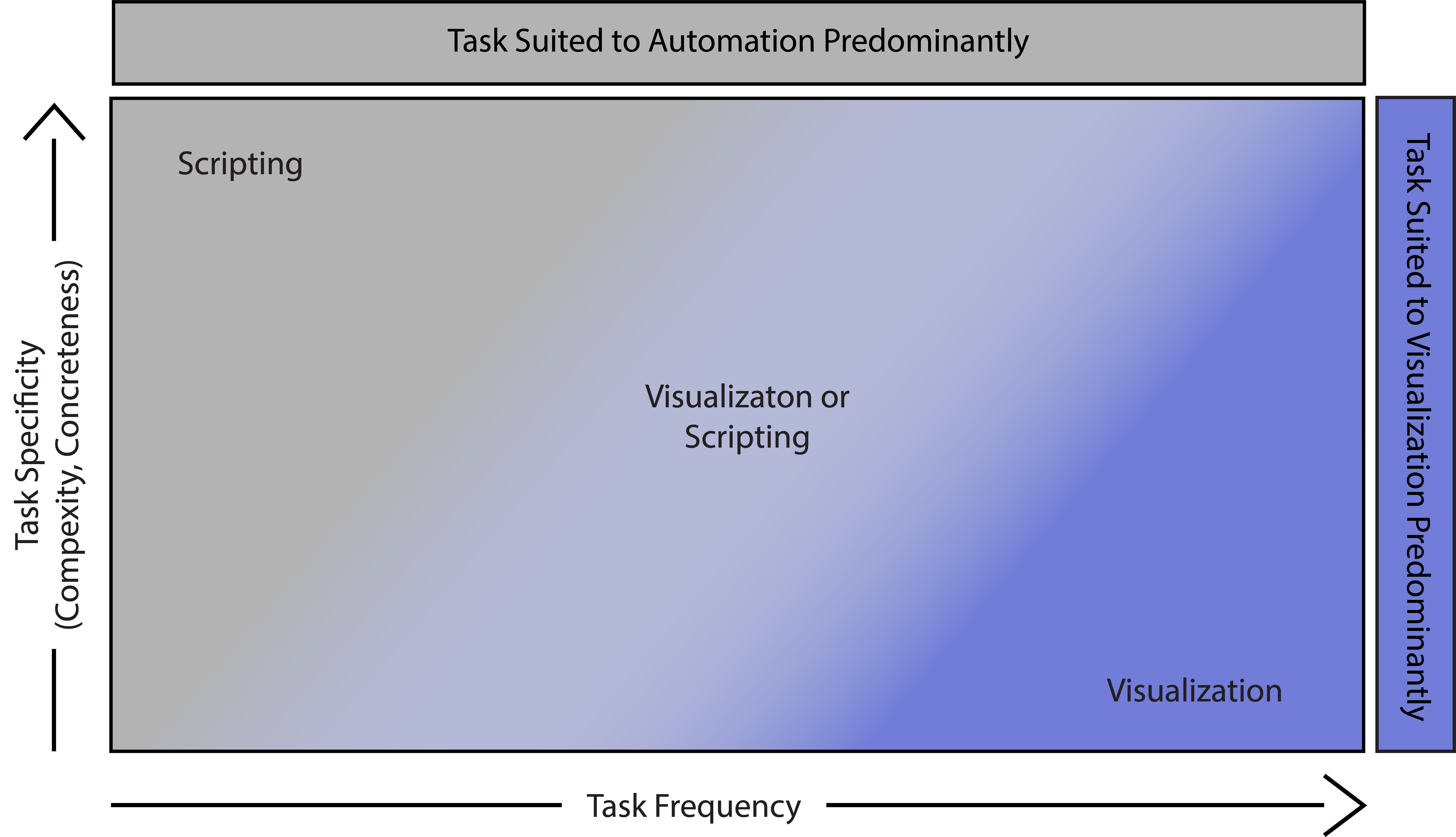}
    \caption{\revision{We considered task frequency and specificity when assigning tasks to scripting or interactive visualization for our design. Highly specific tasks, such as complex queries with precise numbers we assigned to scripting as it offered expressivity and efficiency to our scripting-familiar audience over complex visual interfaces. Less-specific, more frequent tasks like finding anomalies we assigned to visualization as it supports multiple forms of recognition and browsing. We note many tasks can be supported by both, with a hand-off as the analysis grows from more exploratory to more concrete, and thus these assignments reflect a prioritization rather than hard design constraint.}}
    \label{fig:anticipation-complexity}
\end{figure}

\textbf{Suitability} describes how well a task can be supported by visualization or scripting. For most tasks, both are possible. However, some tasks, like following links or making holistic judgements about structure, are much better suited to visualization. Others, like running a predetermined calculation, are much better suited to automation, indicating scripting may be more suitable.

After accounting for suitability, we considered specificity and frequency. These aspects form a space, where more-frequent, less-specific tasks are generally well suited to visualization and more-specific, less-frequent tasks are generally well suited to scripting. \autoref{fig:anticipation-complexity} \revision{illustrates how we conceptualized} suitability, specificity, and frequency in determining which tasks should be supported by the visualization.

\revision{We note these dimensions are not the only factors that could influence whether a task should be prioritized in visualization or assumed to be scripted. For example, how the task is situated in terms of the other tasks in the workflow, facility with scripting, personal preferences towards keyboard interactions, or even transient influences like the user's mood. We chose to prioritize suitability, flexibility, and specificity in our assessment, in part due to some factors being highly variable (e.g., mood) or at the other end of the spectrum, fairly set (e.g., keyboard users in this domain~\cite{isaacs2019ascii}). However, this choice is also a statement of value, similar to choosing a set of design principles, as well as a simplification. More and different factors may also be fruitful to consider, depending on the project.}

\revision{In this case, our collaborators have constrained the workflow to be within the Jupyter notebook and easing the shift between scripting and visual tasks is one of our goals. Thus, we did not prioritize the cost of switching between scripting and visualization contexts as a factor in our assignment. Furthermore, our intended users all have significant programming experience, thus we cast this classification under that assumption. Projects intended for people without such capabilities would assign all tasks to a non-programming interface.}

\revision{An ideal solution might have all tasks supported by multiple modes of interaction, including both scripting and visualization, to account for all factors including personal differences. Our design represents a {\em prioritization} of which tasks to implement in which context, given limited resources and following design study advice of `satisfy, rather than optimize'~\cite{sedlmair2012design}.}

\subsection{Task Analysis}
\label{sec:justtasks}

\begin{figure*}
 \centering
 \includegraphics[width=0.85\linewidth]{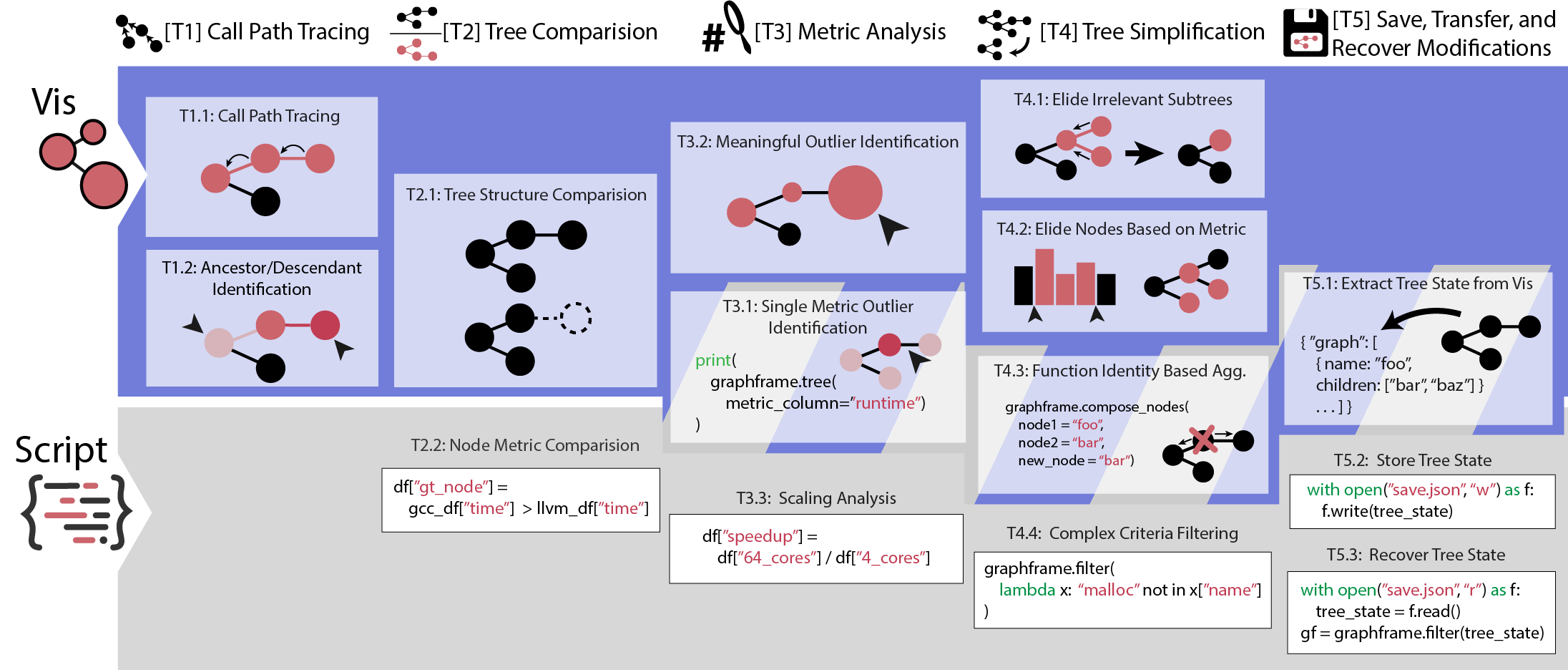}
 \caption{We identify five major tasks which performance analysts engage in when analyzing calling context profiles and classified whether they would be better supported through interactive visualization (purple), scripting in Jupyter (gray), or both. For each subtask we show an example of how it might be accomplished in the identified modality.}
 \label{fig:teaser}
 \end{figure*}

Our task analysis is informed by semi-structured interviews with two Hatchet users, artifacts collected from them, feature requests communicated through the Hatchet project manager, and ongoing meetings with the Hatchet team. The supplemental materials include data supporting each task.

We present our analysis of the tasks including a discussion of their specificity, frequency, and suitability leading to our assignment to visualization (\textbf{V}) or scripting (\textbf{S}) \revision{for the ensuing design iterations}. \revision{The task descriptions refer to the tree task typology of Pandey et al.~\cite{pandey2021treetasks} to abstract the task into general visualization terms of task actions and targets.} \autoref{fig:teaser} shows an overview of tasks, subtasks, and their context which we describe here.

Applying our approach above, we note that in general we determined the visualization should take the lead on structural operations, provided they were not too complex. \revision{These operations tended to either be well-suited to visualization (e.g., involving subtrees) or in the high frequency and less specific quadrant of our space.} We chose scripting \revision{for high specificity contexts like compound queries and other complex operations as well as operations which cannot be supported by the visualization.}

\vspace{0.5ex}
\noindent
\textbf{T1: Call Path Tracing.} \revision{Our interview participants expressed that they wanted} to follow the chain of function calls leading to or from nodes of interest. This provides context to a node of interest, orienting users with respect to code, and allows them to analyze the call stack. \revision{One interviewee noted that ``If you did this filtering you've lost the common parent node of the tree, so it made it very hard to orient yourself about where you were." Another that ``...trees are really helpful, because it gives me basically a call stack. And that makes it much easier for me to understand like where stuff is going."}

\revision{These are \textit{lookup} tasks \cite{pandey2021treetasks}}, either with a path target or more local node ancestor or descendent targets. We assign both to visualization as they are high frequency, low specificity and well-suited to visualization, \revision{especially given the orientation aspects related to these tasks expressed by our interviewees.}

\vspace{0.25ex}
\noindent\textit{Subtasks}
    \begin{enumerate}[topsep=0pt,label=\arabic*.]
        \itemsep=0.25ex
        \item Trace the path from a node to the root. (\textbf{V})
        \item Identify parents and children. (\textbf{V})
    \end{enumerate}

\vspace{0.5ex}
\noindent
\textbf{T2: Tree Comparison.} Performance is often analyzed by comparison between two executions of the same program run with different resources, libraries, or versions. Users want to see how node attributes differ between runs and how the structure of the tree changes between software updates. \revision{We observed these comparisons throughout our discussions where comparative metrics like ``speedup'' and examples with images of multiple trees were shown, including the original Hatchet paper.}

There are two key targets of these \revision{\textit{comparison}~\cite{pandey2021treetasks} tasks,} subtrees and values associated with nodes. \revision{Comparing} subtrees are well-suited to visualization \revision{because it is a structural task}. Node value comparison has high specificity \revision{because often mathematical operations are used to derive new metrics, but relatively low} frequency \revision{as we did not observe more than a few derived metrics created or discussed.} \revision{As it is high specificity and low frequency, we determined it to be more suitable for scripting.}

\vspace{0.25ex}
\noindent\textit{Subtasks}
    \begin{enumerate}[topsep=0pt,label=\arabic*.]
        \itemsep=0.25ex
        \item Compare node values between scaling runs (\textbf{S})
        \item Compare tree structures across implementations (\textbf{V})
    \end{enumerate}

\vspace{0.5ex}
\noindent
\textbf{T3: Metric Analysis.} \revision{Both our interviewees as well as the project manager spoke about understanding the} distribution of attributes across nodes and which nodes have extreme or abnormal attributes. For example, if the speedup metric for a node is unexpectedly small, users may want to investigate it further. \revision{They further} underscored a common need for bivariate analysis: users wanted to know individual metrics like speedup in the context of how long a node executed. A node with poor speedup might not matter much if the node's execution time is already short. \revision{The project manager explained this for one user as, ``He is looking at what is the speed up or slow down right, so if a... function slows down by a lot, the second question he wants to know is, does he care, you know if this is something tiny."}

The definition of ``extreme'' and ``outlier'' is by judgment of the expert, \revision{making these \textit{explore}~\cite{pandey2021treetasks} tasks with node value targets.} This lack of specificity combined with high frequency led us to assign them to visualization. As the definition of ``extreme'' can become more concrete or may be threshholded, making it more specific, we also assigned the subtask to scripting.

Analysis across multiple trees is done by deriving a metric \revision{occasionally, using mathematical operations which can be arbitrarily complex.} Accordingly, due to its relatively infrequent occurrence and \revision{high} complexity \revision{(i.e., specificity), with similar rationale to} [T2.1], we assigned it to scripting.

\vspace{0.25ex}
\noindent\textit{Subtasks}
    \begin{enumerate}[topsep=0pt,label=\arabic*.]
        \itemsep=0.25ex
        \item Find nodes with extreme single metric values (\textbf{S, V})
        \item Find meaningful outliers through bivariate analysis (\textbf{V})
        \item Analyze how node values change across datasets (\textbf{S})
    \end{enumerate}

\vspace{0.5ex}
\noindent
\textbf{T4: Tree Simplification.} \revision{While working with the Hatchet team, we observed datasets with thousands of nodes. Stakeholders uniformly expressed a desire to get to the relevant parts of their trees quickly. They want} to simplify the tree to subtrees of interest for a particular analysis or to elide nodes related to code they cannot change. These nodes may represent details that do not match their level of abstraction in thinking about the code. Users want to aggregate to maintain context without details. The concept of ``interest'' changes depending on the particular analysis. Users can describe a wide variety of queries and filters to retrieve or remove nodes.

\revision{Most tree simplification tasks ultimately support \textit{explore}~\cite{pandey2021treetasks} tasks by removing targets such as subtrees and nodes that are of lesser interest.} Simple elision of subtrees and nodes are low specificity, high frequency, so we assign them to visualization. However, when the criteria for removal becomes complex (highly specific), we assign to scripting. We expect both simple and complex criteria.

Aggregation differs in that it changes the tree structure. Structure tasks are typically well-suited to visualization, but some of the requests were complex and specific, e.g., ``elide paths matching A then B then C.'' Thus we assigned scripting and visualization to cover both simple and complex cases.

\vspace{0.25ex}
\noindent\textit{Subtasks}
    \begin{enumerate}[topsep=0pt,label=\arabic*.]
        \itemsep=0.25ex
        \item Elide subtrees not relevant to current analysis (\textbf{V})
        \item Elide nodes based on a metric (\textbf{V})
        \item Aggregate subtrees or internal paths based on structure and function identity (\textbf{V, S})
        \item Filter based on some other complex criteria. For example, ``Remove all nodes in a specific library'' (\textbf{S})
    \end{enumerate}

\vspace{0.5ex}
\noindent
\textbf{T5: Save, Transfer, \& Recover Modifications.} \revision{Across our interviews and discussions with the project manager, we heard a desire to stop and re-start their analysis across sessions as well as to share their notebooks with others. Thus, they need} to reproduce the modifications they made to the tree, for use across different analysis sessions or when sharing with others. They may make changes either visually or through scripting and then transfer those changes between the two contexts.

\revision{These are not direct tree analysis tasks in the scope of Pandey et al. but echo the \textit{extract} task proposed by Shneiderman~\cite{ShneidermanEyes}} and are required for an \revision{integrated} scripting and visualization workflow. They are well-suited to scripting by the nature of loading a notebook, but also very specific and relatively infrequent in exploration. The first subtask is split between visualization and scripting because its tie to the visualization makes it highly visualization-suited.

\vspace{0.25ex}
\noindent\textit{Subtasks}
    \begin{enumerate}[topsep=0pt,label=\arabic*.]
        \itemsep=0.25ex
        \item Extracting the tree state from the visualization (\textbf{V, S})
        \item Store the tree state for future use (\textbf{S})
        \item Recover the tree state (\textbf{S})
    \end{enumerate}

\section{Scripting-Embedded Visualization Design}
\label{sec:cctvis}

\begin{figure*}
    \centering
    \includegraphics[width=0.95\textwidth]{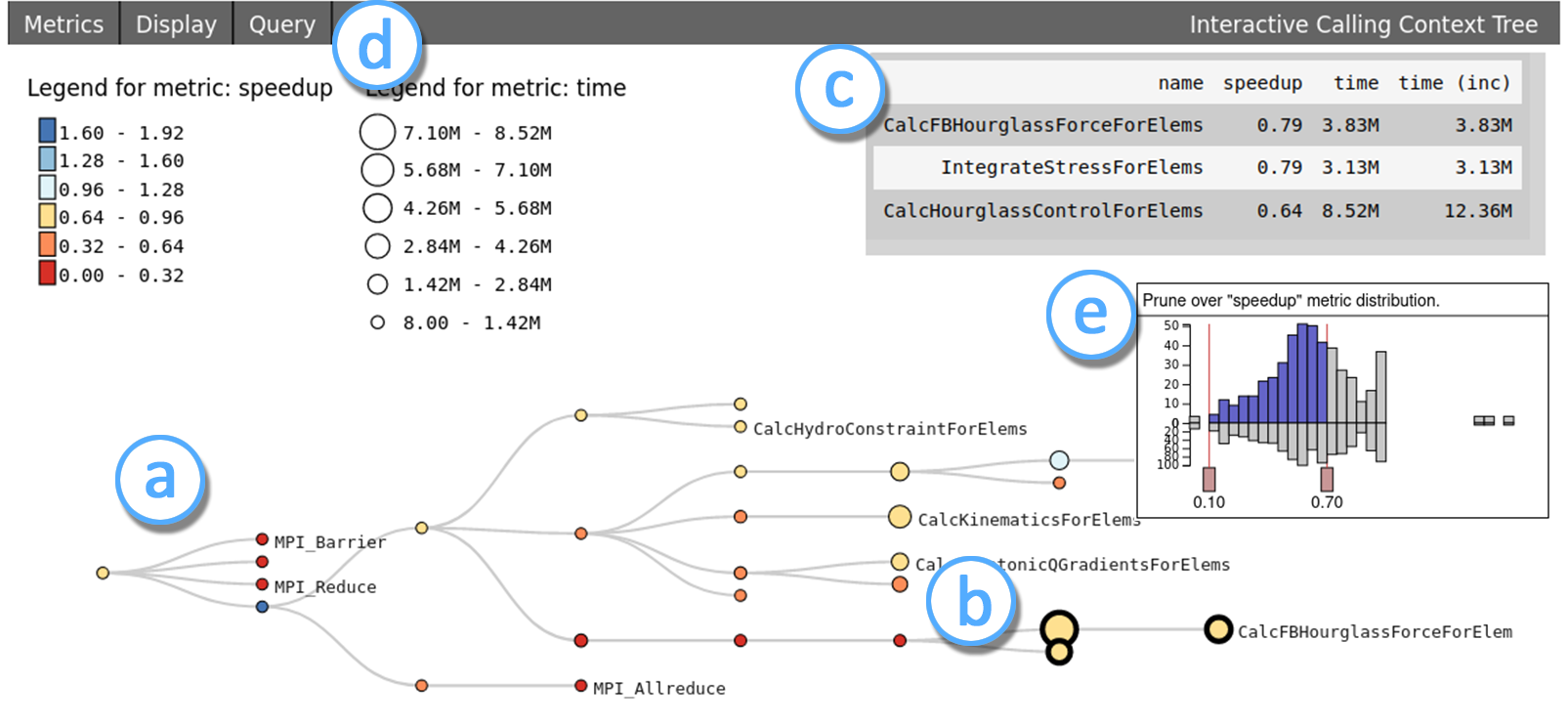}
    \caption{Embedded CCT Visualization. The main view (a) is a pannable, zoomable node-link diagram with two node encodings---color and size. Selected nodes (b) have a thick border. Their details are shown in a floating table (c). Features like mass-pruning (e), changing encoding metrics, and exporting queries are available through the menus (d). }
    \label{fig:vislayout}
\end{figure*}

We designed the visualization through an iterative process, sharing design documents and prototypes in weekly meetings. Collectively, interviews with frontline analysts, feedback from the Hatchet team, and our resulting task analysis informed our design. In addition to these design inputs, there were two other constraints we sought to fulfill: {\bf [C1]}: the visualization would be embedded in Jupyter notebooks as this is an intended environment by the Hatchet team, and {\bf [C2]}: the visualization needed to be intuitive as the Hatchet project manager had observed target users would not take time to learn an unfamiliar visualization. 
We explain the key elements of our interactive, embedded CCT visualization (\autoref{fig:vislayout}). 

\subsection{Tree View and Bivariate Encoding}
\label{sec:maintree}

We chose a node-link depiction as the main representation for multiple reasons. First, it was familiar to our users and prevalently used in this space, decreasing the learning barrier to use and supporting our intuitive constraint, C2. Our discussions with collaborators indicated they view node-link depictions with parents centered over children as {\em true} trees, while \revision{other depictions,} even indented trees with explicit orthogonal links \revision{where parents exist at one end of the list of nodes}, like Hatchet's built-in drawing {\em and those of file browsers}, were not considered real ``trees."  

We considered Sankey diagrams, as had been used in CCT visualizations like CallFlow~\cite{kesavan2021scalable}, and included images from CallFlow along with hybrid approaches in low fidelity mock-ups, but the node-link diagram was preferred.

We did not pursue adjacency-based idioms like icicle plots and sunbursts because their design implies the attribute value of a parent node is the sum of the attribute value's of its children. This property does not hold across common metrics in the analysis and in several cases, child attribute values can be larger than that of their parent node.

We chose to orient the node-link diagram expanding rightwards to make use of Jupyter cell aspect ratios. Users can pan and zoom the tree with mouse/trackpad interactions. 

To support the metric analysis tasks T4, we encode two metrics, a primary metric with node color and a secondary metric with node size. Users can change the metrics as well as switch between a diverging color map and a single hue ramp, both which can be inverted, through the menus. Legends for both encodings appear under the menu bar.

Bivariate support was a priority feature as often metrics are evaluated in the context of execution time. Early designs used color for a single metric, due to its distinguishability and preattentive nature to support outlier detection. Color is often used in other HPC visualization~\cite{adhianto2010hpctoolkit, malony2014performance}, so we expected familiarity from our users as well. We considered a bivariate colormap to put metrics on equal footing, but ultimately chose size as the second channel so metrics could be decoded individually, stand out individually, and intersections would still be salient. 
 
 Users can access additional attributes through node selection. Details of selected nodes are shown in a floating table, designed to look similar to the pandas output. Nodes may be selected on click or by brush.

\subsection{Tree Simplification (Pruning)}
\label{sec:pruning}

Calling context trees can contain thousands of nodes, not all of which are of interest to a given analysis. However, analysts want access to all of them should their analysis require it and also for initial overviews. Thus, supporting the tree simplification tasks (T4), we added several interactive features for reducing the size of the tree shown.

We note these tree elision functionalities are {\em pruning} operations and not filtering because they only elide nodes from the leaves towards the root (i.e., subtrees) to maintain the context of a connected tree structure. 
We provide a manual prune (T4.1, subtree elision) and a mass prune (T4.2, elision based on metrics). Elided subtrees are signified with a black arrowhead, indicating that they can be expanded. 

\begin{figure}
   \centering
   \includegraphics[width=\columnwidth]{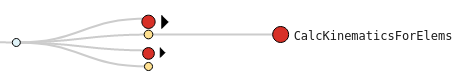}
   \caption{\revision{Manually pruned subtrees are depicted with arrows indicating they can be expanded on click. The color and size encode the average value of associated metrics in the elided subtree.}}
   \label{fig:pruning}
\end{figure}

With {\bf Manual Pruning}, double-clicking a node collapses (or uncollapses) its subtree \revision{(\autoref{fig:pruning})}. {\bf Mass Pruning} elides subtrees based on a metric value range, set through a floating interface (\autoref{fig:vislayout}(e)) which is available through the menus. The interface shows the distribution of the primary metric as a butterfly histogram. The top histogram shows the distribution of prunable nodes. The bottom shows the distribution of nodes that are internal, i.e., they are not pruned because doing so would disconnect the tree.

The pink handles can be adjusted to set the range shown. On launch, this is set to elide nodes with 0.0 as their primary metric value to limit the number of empty subtrees drawn on initial load and enhance the scalability of this visualization.

\begin{figure*}[htp]
    \centering
    \includegraphics[width=\linewidth]{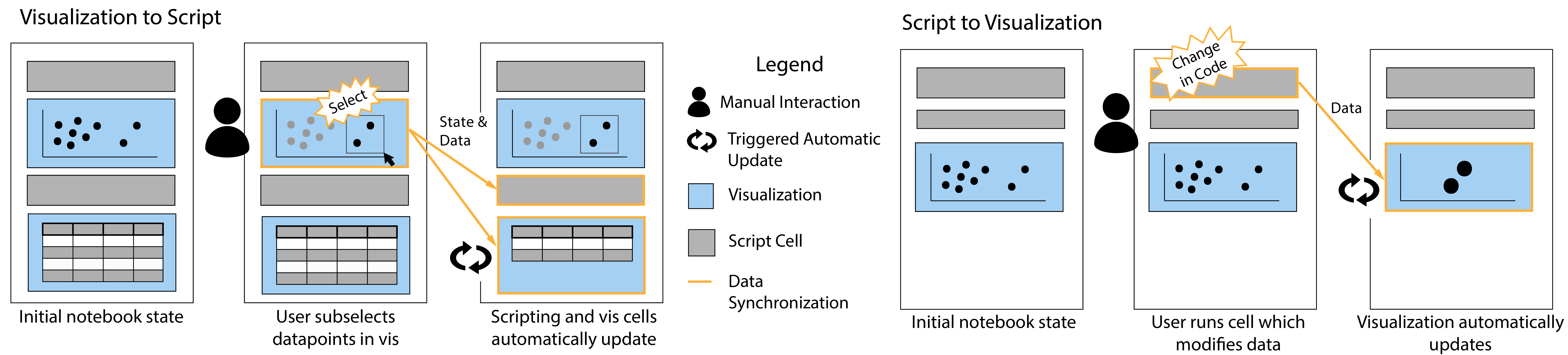}
    \caption{We illustrate two scenarios, each showing one direction in the two-way visualization+scripting paradigm. Each row of captioned figures shows a Jupyter notebook over time.  In the first, "Visualization to Script," the user makes a selection in the visualization, causing automatic updates in cells that use the visualization state: a scripting cell and another visualization cell. In the second scenario, the user changes data stored in a variable and re-runs a code cell. A visualization showing the modified variable automatically updates to reflect the state of the notebook.}
    \label{fig:script-vis-explination}
\end{figure*}

\begin{figure}[ht]
    \centering
    \includegraphics[width=\columnwidth]{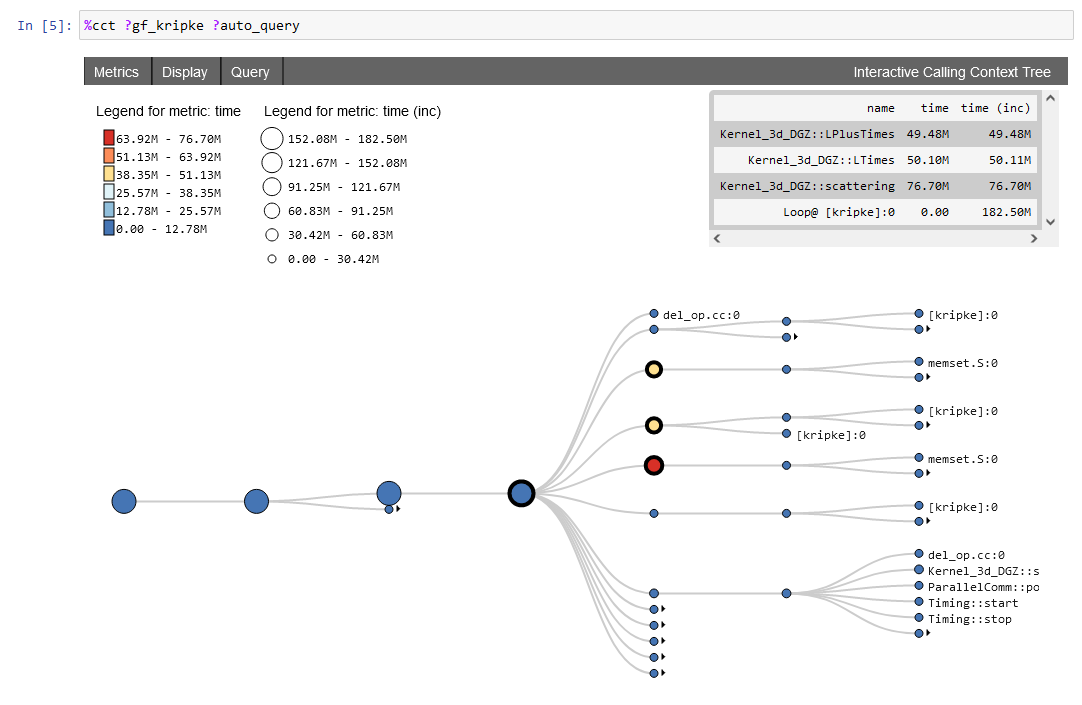}
    \includegraphics[width=\columnwidth]{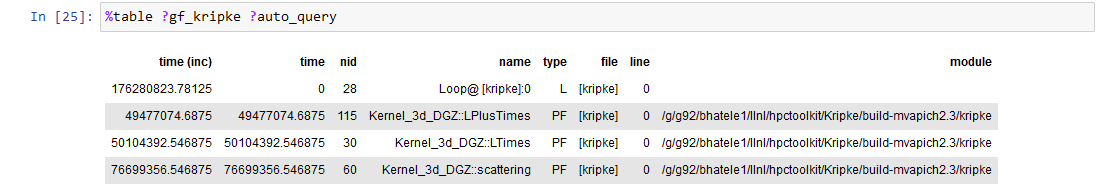}
    \caption{An example of a linked table view. The table in the bottom half of the figure is a distinct view running in a different cell from our CCT visualization. It automatically updates to reflect selections made in the CCT, showing additional information about each node.}
    \label{fig:table_view}
\end{figure}

Outside of tree simplification, further optimizations were undertaken to enhance scalability by mitigating label clutter. We describe this in our supplemental material.

\subsection{Visualization \& Scripting Workflow Design}
\label{sec:embeddeddesign}

To support those tasks which require communication between visualization and scripts, we embed custom visualizations into the notebook and provide mechanisms for transparent data flow between visualization and notebook.

\subsubsection{Data Flow between Visualization and Notebook}
To move scripting results to the visualization, users  associate a Hatchet GraphFrame with the visualization's Jupyter magic function. Users can flexibly derive new data via Hatchet or general Python scripting and then visualize. We expect users to implement more complex and hard-to-predict filtering tasks (T4.4) and precise metric comparisons and searches (T2.1, T3.1, T3.3) this way.

Conversely, we pass data and state information from the visualization back to the scripting cells by exporting a query in the Hatchet query language. This supports our T5 goals of saving and recovering tree state as users can apply the output query to a GraphFrame to retrieve the pruned tree of interest. We chose to export selections in the Hatchet query language rather than as Python queries following collaborator requests. They wanted users to be able to relate the query back to the tree-structure using a graph-friendly, purpose-built syntax.

\subsubsection{Manual and Automatic Updating Interaction Modalities}
\label{sec:updatemodalities}

\revision{To lower the temporal gap identified by Wu et al.~\cite{b2}, we proposed a design where groups of cells would automatically update with data changes. We hypothesized such a design, where multiple perspectives have linked updating, might have advantages to multi-view visualizations, despite Wu et al.'s identified layout gap. However, we anticipated the choice of which cells should be strongly linked was hard to predict. Furthermore, the Hatchet project manager was concerned that linked updating might confuse users. Thus, we implemented two modalities, an automatic one where changes to one cell cause others to reload and reflect changes, and a manual one where each cell must be executed like normal.}

\revision{Specifically,} in the manual modality cells must be run explicitly by the user and saving state \revision{is} done via the visualization. Thus, filters applied in scripting require re-running the visualization. 

In automatic modality, we can mark cells to automatically update when specific data changes either in the scripting cells or the visualization. Thus, filters applied in scripting can automatically trigger a visualization change and the visualization state can be available in scripting cells continuously. This also allows the creation of linked views across cells.

\autoref{fig:script-vis-explination} illustrates two automatic update scenarios, one where the user updates the visualization and one where the user updates the script. In the first, data and state flow from a visualization back into a notebook. When a user interacts with the visualization, like selecting points in a chart, the selected data and state information is automatically updated in the notebook for use in subsequent cells with no need to run a manual retrieval function. Furthermore when this data/state information is provided as an input to another visualization, it will also update to reflect these changes. 

In the second scenario, a change to the data in a scripting cell will automatically trigger an update in an visualization which uses the modified data as an input parameter. With this automatic updating, users can focus primarily on tweaking scripts and modifying their data without pausing to intentionally re-run and sync associated visualizations. 

Users can define which data and cells invoke automatic updating using the question mark operator from the Roundtrip library~\cite{Roundtrip}. In particular, data marked for automatic updating is preceded by a question mark, indicating it is being watched in that cell. The visualization is similarly bound as shown below with its data (``GraphFrame'') and its state in terms of pruning and selection (``VisStateQueries''):
\begin{lstlisting}
%cct ?GraphFrame ?VisStateQueries
\end{lstlisting}

\noindent
Here \texttt{\%cct} invokes our visualization.

\autoref{fig:table_view} shows a linked table view. With the use of watched variables and state passing, the table generated in the second cell will automatically update based on the nodes selected in the tree visualization. We use this example in our follow-up study to explore this paradigm.

\section{Visualization Evaluation}
\label{sec:eval}

We conducted two studies. The first was to validate that our visualization design and integrated scripting workflow supported the identified tasks and more broadly how it addressed the needs of people who analyze performance profiles.

In Section~\ref{sec:study} we follow up with the same experts to more deeply explore the scripting-visualization workflow and user preferences of interaction modalities.

We summarize the initial study due to space constraints, focusing on the elements that provide context for the second study and the findings related to the visualization and scripting design. A more detailed description of the study, results, and findings is in the supplemental materials.

\subsection{Evaluation Design}
\label{sec:evaldesign}

We conducted \revision{one hour} sessions via video-conferencing. Each session consisted of an initial briefing, overview of visualization features (25 minutes), three tasks for the users to complete (15-20 minutes), semi-structured interview questions (10-15 minutes), and a debriefing. 

We limited the evaluation to the Manual Updating interaction mode due to concerns of the Hatchet project manager, who is also an experienced HPC performance analyst. They thought users would find Automatic Updating mysterious and might accidentally overwrite state they wanted to save. They argued the extra operation was small in terms of the entire analysis. We address Automatic Updating mode in the follow-up study (Section~\ref{sec:study}).

\inlinehdr{Evaluation Datasets.} 
For the tutorial, we used a small dataset collected during an execution of Lulesh~\cite{LULESH}. For the evaluation tasks, we used larger datasets collected from two Kripke~\cite{kripke} executions, run on 64 and 128 cores respectively, resulting in trees of 1500 and 2700 nodes. 

\inlinehdr{Participants.} 
We recruited seven participants (\autoref{tab:participants}). P1 and P2 were the front-line analysts that we interviewed early in the design study. P3, P4, and P5 were all professionals at the same organization but in different software teams. P6 and P7 were PhD students in HPC. Only P4 said they were familiar with Kripke. P1 stated they were aware of it, but not familiar.

P1, P2, and P5 are active Hatchet users, P3, P4, and P6 have some familiarity, and P7 had not used Hatchet before. P5 is an active Jupyter user. P1, P2, P4, and P6 used Jupyter regularly but needed reminders of how to expand and run cells. P3 and P7 had only cursory knowledge of Jupyter. 

\begin{table}[h]
    \centering
    \caption{Participant Familiarity with Hatchet and Jupyter}
    \scriptsize
    \label{tab:participants}
\begin{tabular}{ c | c | c | c | c | c | c | c}
 & P1 & P2 & P3 & P4 & P5 & P6 & P7 \\
 \hline
 Hatchet & High & High & Some & Some & High & Some & None  \\  
 Jupyter & Some & Some & Little & Some & High & Some & Little     
\end{tabular}
\end{table}

\inlinehdr{Evaluation Tasks.} We list evaluation tasks and their corresponding task analysis items below. The first task asked participants to familiarize themselves with the dataset and enabled us to observe insights generated in open exploration. The second task gave participants a common high-level domain task so we could observe if and how the design served it.
The final task focused on having the participants try the visualization-to-scripting half of the workflow.

\begin{enumerate}[label=E\arabic*.]
    \itemsep=0.25ex
        \item Open Ended Exploration [T1.1, T1.2, T3.1, T4.2, T4.3]
        \item Identify A Candidate for Optimization [T2.1, T3.1, T3.2]
        \item Export and Apply Query [T5.1, T5.3]
\end{enumerate}

\subsection{Summary of Evaluation Analysis}
\label{sec:evalanalysis}

Descriptions of participant activities and our hierarchically organized themes and codes are available in the supplemental materials. We summarize findings related to the validation and a more detailed discussion of the notebook-embedded workflow findings.

\inlinehdr{Validation Summary: } 
Participants successfully completed every task they were asked, except for P5 on E2 who uttered several insights, some of which other participants gave as their answers, but said they would need more time to come up with a better answer. We also observed participants vocalize several other meaningful insights \revision{about the data} during the tasks.

Three participants said the visualization had everything they needed. Favored features included the mass prune, bivariate encoding, and ability to compare trees. Each feature was mentioned at least once. Five participants expressed finding the interactions with the visualization intuitive. Three participants expressed a desire to work with this visualization further.

Requests included undo features (P1-P4, P6), pruning by brushing (P4, P7), changing the histogram range (P2, P5), the ability to add annotations (P4, P7), and several one-off quality-of-life issues. There were also a variety of requests for other metrics or to mass prune by other criteria.

\inlinehdr{Notebook-Embedded Workflow.} 
\label{sec:evalembedded}
We observed that most participants worked through the tasks which required scripting or switching between code and visualization contexts without significant frustration or friction, though we note these opportunities were limited by the length of the evaluation. In particular, while all participants suggested a complex query that we made the conscious choice to be handled by scripting, only one (P5) was prepared to implement it.

The complex queries suggested by the participants were varied. The most prevalent was to prune calls related to MPI, a specific domain standard (P2, P3, P4, P5). This operation cannot be done with a simple name match and may not be useful for other datasets. Individual suggestions included pruning framework-related calls, system calls,  all libraries, libraries qualified on metrics, ``what we control,'' arbitrary regular expressions, pruning based on depth, and pruning leaves. This wide variety of desired, complicated queries suggests the need for the flexibility inherent to scripting. 

We probed participants further about whether their query should be supported by scripting or visualization. P2 and P5 chose scripting, with P5 explaining that debugging regular expressions would be easier. P4 expressed an interest in both, but prioritized scripting support before interactive visualization. 

Three participants (P3, P5, P6) inquired about the possibility of adding metrics to the dataset while working on the speedup task. P5 suggested ``percentage of time,'' which is something Hatchet can do in a few lines of code. Adding metrics is another task supported by the integrated scripting environment.

Not all participants were comfortable with the embedded approach. P3 described the workflow as ``not helpful.'' P1 said they think of scripting and visualization as separate, so the workflow was ``unintuitive,'' but qualified their comment, noting the workflow has a need, but they were unsure how to implement it. 
Most participants did not comment on the speedup calculation or the query saving, suggesting the interface was not particularly notable to them.

P1, P4, and P5 liked the capability of retrieving a query from the visualization both for saving and retrieving the subset, but for other reasons as well. P1 was interested in retrieving the subselection as a table for further analysis. P5 said they did not have a personal use, but ``could see [their] colleagues using this to build queries.''

\subsection{Limitations and Threats to Validity}
\label{sec:threats}

The generalizability of our evaluation is limited by the modest number of participants. We sought participants who would be representative users---those \revision{with HPC performance analysis and Hatchet experience}. This limited our participant pool. 

All of the participants knew at least one of the authors, which may have biased their comments towards expressing positive sentiments, e.g., P2 declined  to compare our visualization to a commercial tool.

Though Jupyter is a delivery mechanism for Hatchet and other tools in our collaborator's organization, familiarity with Jupyter varied, with only one participant being a frequent user. The lack of familiarity may have affected feedback and use. 

One hour is a typical schedule slot duration for our professional users, which limited the kinds of tasks we could include. Even though most participants had some familiarity with Hatchet, they were not so familiar with the API that we could expect them to write Hatchet queries in open-ended analysis on the fly, though one, P5, did.

\section{Visualization-Scripting Study}
\label{sec:study}

\revision{As discussed in Section~\ref{sec:updatemodalities}, we implemented an automatic updating modality to lower the temporal and layout gaps expressed by Wu et al. and make the design more similar to multi-view visualizations. However, the Hatchet project manager was concerned the users might find this confusing. We thus ran a follow-up study, six months after the validation one, focusing more on the Manual and Automatic Updating modes than the visual design. We wanted to understand if focus between scripting and visualization would become more frequent if barriers were lowered and how the modes would be perceived by users.}

\subsection{Study Design}
We conducted 60 minute sessions via video-conferencing. Each session consisted of a briefing, overview of visualization features (25 minutes), three tasks for the users to complete (15-20 minutes), a semi-structured interview (10-15 minutes), and a debriefing. 
The tutorial refreshed participants on the visualization from the initial study and taught them about the Automatic Updating modality. The tasks and interview questions focused more on interaction than achieving a goal.

We added pre-written programmatic filter templates to the working notebook so participants could adjust and apply them to the dataset. This choice was based on our observations in the initial study that most users did not have time to fully construct their own. We included a string filter, a value filter, and a tree depth filter.

\inlinehdr{Participants.} We invited the participants from Section~\ref{sec:eval} to leverage their familiarity with the visual design, allowing us to focus on the interactions. All began the study, but P7 had to leave due to technical difficulties before the tutorial finished.

\inlinehdr{Study Datasets.}
We used the 64-core Kripke dataset from Section~\ref{sec:eval}. We chose to use the same dataset because we knew from the prior study it was well-sized and interesting enough for our study time constraints. We were not concerned about familiarity because we were seeking observations regarding different interaction modalities, not performance insights to validate the visual design. In running the study, we found participants did not recall much over the six month gap.

\inlinehdr{Study Tasks:}
Our tasks focused on interacting with the visualization in different modalities.

\begin{enumerate}[label=E\arabic*.]
    \itemsep=0ex
        \item Filter \revision{the CCT to a manageable size (automatic updates)}
        \item Filter \revision{the CCT to a manageable size (manual updates)}
        \item Identify the line number and file associated with a call site using automatic-update and the linked table view.
\end{enumerate}

E1 and E2 explore the scripting-to-visualization half of the workflow and ensure participants have tried both \revision{automatic and manual modalities} before the interview. \revision{Based on the previous study, we anticipated they would exercise multiple filters to winnow the CCT to data that interested them.} E3 explores the visualization-to-scripting side. We alternated the order of E1 and E2 between participants.

\inlinehdr{Hypotheses.} We had two hypotheses:
\begin{enumerate}[label=H\arabic*.]
    \itemsep=0ex
        \item \textbf{Participants will iterate more on their filters in Automatic Updating mode.} We expected the more fluid workflow would be less discouraging to altering filters.
        \item \textbf{Participants preferences between modes will be mixed.} We based this hypothesis on concerns of P1, P3, and the Hatchet project manager in the validation study.
\end{enumerate}

\inlinehdr{Semi-structured interview.} To elicit discussion about the notebook-embedded workflow, we asked:
    
\begin{enumerate}[label=Q\arabic*.]
    \itemsep=0ex
        \item Which mode do you prefer? Why?
        \item Would you prefer pure visual interaction to these?
        \item Do you have any feedback on selecting in the visualization and using selected data in subsequent cells?
        \item Do you have other thoughts on this auto-updating functionality?
        \item Do you have any other suggestions?
\end{enumerate}

\subsection{Evaluation Task Results}
We summarize our observations during the evaluation tasks.

\inlinehdr{E1 and E2: Filter a dataset.}
    Participants were asked to filter the dataset to something that felt manageable. P1 and P4 started with the Manual Updating modality, where each cell must be run explicitly when data changes. P2, P3, P5 and P6 started with Automatic Updating, where data changes in any cell can cause others, including the visualization, to automatically update. 
    
    Participants used a variety of strategies. All but P6 used graphical interactions in the visualization and programmatic filters across both tasks. (P6 used the mass prune only.) In Automatic Updating mode, P2 and P5 used programmatic filters only.

    \revision{All participants used multiple distinct interactions and/or filters in E1.} String filtering was the most common template used. P1-P4 quickly identified and removed large subtrees related to specific libraries. P5 used only the metric value filter. P1 and P3 followed their string filtering with the mass prune. P4 used the value filter after the string filter. 
    
    \revision{In E2,} P2 and P4 \revision{re-applied} the programmatic filters they built in \revision{E1} to the refreshed dataset\revision{, thus iterating \textbf{less}}. P1, P3, P5, and P6 used the mass pruning in E2 and exported the tree state to synchronize the data with the visualization. 

\inlinehdr{E3: Identify Additional Information.} 
    \revision{We guided users to run a cell with a linked table view we developed for this study (\autoref{fig:table_view}) using Automatic Updating.} We asked them to select a node and identify the line number and file it was associated with. Due to timing issues, only P1-P3, and P5 were asked to execute this task. They were all able to complete the task by reading the table.

\inlinehdr{State Tracking Observations.}
    Across all evaluation tasks, we observed participants had difficulty tracking the \revision{data state} across the Jupyter cells. P1 was unsure in which cell to run the programmatic filters and seemed not to understand what was running when some cells took a couple seconds to load. They noted that Jupyter is non-linear and running a cell can affect a non-adjacent one. P6 said they could lose track of state without more explicit marking and was unsure of the order some operations took place.  
    
    Several participants (P2-P4) were surprised when a filter they ran did not change the visualization.
    P2 and P3 experienced this error when changing to manual mode and attempting to apply their exported queries to the dataset. This may have been the result of a (mis-applied) learning effect. P4 experienced the error when they failed to adjust what dataset they applied an exported query to.

\subsection{Analysis}
\label{sec:study-analysis}

Two authors who attended or watched the evaluation sessions coded their resulting notes from the videos and transcripts. They then met to derive themes.

\inlinehdr{No express correlation between iteration and Automatic Updating.}
H1, which hypothesized there would be more iteration in Automatic Updating, was not observed. All participants iterated more in the first task because they could apply their knowledge in the second. We did not notice a clear difference in the first task between participants who were assigned Manual Updating and those assigned Automatic Updating. If a difference exists, it is not detectable in this study. 

\inlinehdr{Preference for automatic updating.}
H2, which hypothesized there would be mixed preferences between Automatic and Manual Updating, did not hold. In the interview section, P1, P2, and P4-P6 quickly expressed their preference for Automatic Updating. P3 stated ``they both have advantages,'' but on further probing, seemed to be discussing graphical versus scripting differences. 

P2, P4, and P6 \revision{attributed} their preference to Automatic Updating \revision{to} having fewer clicks. P2 said it ``helped [them] understand data better without getting distracted." P4 noted they did not need to ``look at'' the visualization before applying it, so would not want to manually re-run the visualization cell. P6 stated in addition to the filtering, they \revision{could} extract the state of the visualization without going through the visualization UI, saying it was ``easier to use.''

P1 described Automatic Updating as ``better for clicking and exploring stuff,'' but noted it runs the risks of users getting ``lost" in the state. In contrast, P5 found the Automatic Updating behavior ``more obvious'' and said it removes state tracking burden because they can set it up and not worry about keeping track of which cells to re-run, especially because they could be out of order.

\inlinehdr{Preference for visualization-scripting workflow.}
    All participants, expressed a preference for a visualization-scripting solution over an exclusively point-and-click visualization. P1-P4 directly described the visualization as intuitive for initial exploration but then wanted it paired with scripting for handling specific cases (P1, P3), documenting, saving, and sharing (P1, P2, P4), and applying the useful direct manipulation operations, once discovered, more globally (P2). P1 made an analogy to VisIt~\cite{childs2012visit}, noting how ``90\%'' of their interactions are direct manipulation but having Python available for the others is a boon and that furthermore, they can save their operation history with it.
    
    Without mentioning the visualization directly, P6 echoed sentiments regarding the flexibility of scripting, especially for multi-constraint queries. P5 said they liked doing queries by scripting because they know how to debug Python via small test cases in a notebook, whereas it is not obvious how to debug queries in a visualization interface.

    \revision{We note P1's point-of-view shifted in comparison to the previous study where they expressed reservations and difficulty in imagining how it would work. We cannot claim the introduction of automatic updating in this study helped them conceptualize the workflow however, as other factors such as time or the interview questions may have led them to consider it differently.}

\inlinehdr{Mostly positive sentiments towards linked table view.}
    Of the participant who did E3, finding a file and line numbers in a linked table view (P1-3, P5), all but P2 expressed positive sentiments. P1 found the task well in-line with their existing needs and liked being able to use data later in the notebook. P5 stated ``Oh wow, that's very cool, that's amazing'' and posited a use case where they could select from the tree and then export the data for communicating with colleagues. P2 however said they were ``uncertain'' about the linked view and suggested the visualization could have larger tooltips with the same data.
    
\inlinehdr{Variety of Strategies.}
As in the validation study, we observed a variety of strategies used by participants in manipulating the tree visualization. In addition to the variety of visual interactions first observed, participants each had a different workflow in applying programmatic filters (if any), in what strings and values they specifically used, and in where they applied programmatic filters in their tree manipulation process.

\subsection{Limitations and Threats to Validity}

    This study has the same limitations and threats as in Section~\ref{sec:eval} in terms of population, afforded time, familiarity with Jupyter, and structure. Additionally, there was imbalance in the number of participants starting using Automatic Updating versus Manual Updating due to technical difficulties causing P7 to drop out. 

    We used either a shared HPC system or virtual machine for our evaluations, which led to slowdowns requiring live troubleshooting. Some operations took 1-2 seconds. Both the response times and troubleshooting may have influenced participant responses.

    \revision{Our study is a qualitative, preliminary investigation into trade-offs between automatic and manual updating designs connecting cells in a notebook environment. As linked cell updating is still new to users, observations may not capture how people would respond if such designs were more common. Had we more time per participant and the knowledge gained in this study, we would probe participants for their understanding and expectations of notebook state during tasks as well as how they would group cells. The latter may provide insight into which cells should be more tightly linked. Eye-tracking could provide additional data regarding how participants consult cells which could also inform design.}
\section{Reflections and Design Implications}
\label{sec:reflections}

The embedding of our visualization in an exploratory scripting workflow was an intentional and informed decision, based on the constraints of our collaborators---using notebooks as a delivery mechanism---as well as our observations that performance analysis is an inherently exploratory process that can benefit from the flexibility of scripting. We designed the visualization for an exploratory workflow of modifying the dataset with Hatchet scripts, doing visual tasks, and then either modifying the dataset again or exporting a subset of it found during visual exploration. We discuss our lessons learned regarding our notebook-embedded design process.

\inlinehdr{Considering task specificity and frequency helps match tasks to workflow modalities that support them.} 
\revision{Of the possible factors influencing whether a task is better served by scripting or visualization, we used} task specificity, frequency, and suitability as described in Section~\ref{sec:taskapproach} \revision{to make our determinations}. Feedback from participants largely validated our task assignment choices, though we found evidence for having more tasks supported by both modalities. Even in these cases, participants' prioritizations followed our task analysis.  

For example, high frequency, low specificity tasks in this project included removing ``irrelevant'' library nodes or highlighting ``interesting'' nodes. As ``irrelevant'' and ``interesting'' were contextual and ill-defined at first, these tasks suggested visualization. As they became more specific and thus individually less frequent, they moved to scripting, e.g., remove ``specific libraries'' or ``nodes by metric value.'' \textbf{This evolution suggests designing with both visualization and scripting in mind to support users as task concreteness changes.} Furthermore, as we observed a diversity of requested queries with little frequency among them, by assigning them to scripting, we avoided adding complex \revision{one-off} features.

We observed concreteness as a facet of specificity during P1's reference to the ``replay'' functionality of the scientific visualization suite VisIt~\cite{childs2012visit}, which enables users to save their exploratory processes as code. Although their interactions were originally exploratory, they become more concrete as the user better understood their dataset. It was thus helpful to have them available via script rather than direct manipulation. The Hatchet team corroborated this expected workflow mode. To them, analysts primarily use a visualization to explore and understand their data with the explicit goal of specifying their desired transformations for script-based analytics.  

\revision{There are numerous factors that may influence the appropriateness of assigning a task to scripting or visualization. We chose to prioritize three which worked well in our project. Several other possibly influential factors, such as facility with scripting and where various tasks were situated, were fixed in this case, thereby allowed us to decrease the space of factors considered, but may need to be more carefully considered in other projects.}

\inlinehdr{Consider designing for two-way data and state transfer between scripting and interactive visualization.} In our second study, there was unanimous support for having a scripting interface to the visualization. Benefits expressed by participants included the ability to create complex queries and handle edge cases, share analyses with others, operationalize repeated visual queries, and debug queries. At the same time, participants expressed interest in ease-of-use exploration, with one expressing a 90/10 split between visual interaction and scripting, respectively and most participants preferring Automatic Updating in the exploratory scenarios. Although effective in this case, this approach would be appropriate only for a users who know how to script.

\revision{\inlinehdr{For scripting-comfortable users, we found lookup, explore, and structure-based tree tasks were still best-suited for visualization.} During our task analysis (Section~\ref{sec:taskapproach}), we assigned abstract lookup and explore tasks as defined by Pandey et al.~\cite{pandey2021state} to visualization for their high frequency and low specificity. Additionally, we determined structure-based tree tasks such as comparison and subtree-elision were visualization tasks based on suitability. In contrast, tasks requiring specificity were complex and compound queries used for filtering. We expect these correspondences between these abstract tasks and assignments may be applicable beyond performance analysis when tasks exist in a similar combined environment (e.g., notebooks) and the audience has facility with scripting.}

\inlinehdr{Users need more explicit state tracking when visualization and scripting is tightly bound.} We observed tension between perceived ease-of-use of the workflow and difficulties in tracking state. Participants expressed near unanimous approval of the Automatic Updating modality, elaborating that it saved them effort and noting they found it easier to track state. However, we also observed participants having difficulty in tracking the state of the data object in terms of filters applied or knowing in which order cells would or should be run. Automatic Updating exacerbated confusion \revision{about} which cells were running when updates were not instantaneous.

These issues suggest that this visualization, future embedded visualizations, and computational notebooks in general, could benefit from interfaces that make both the state of the notebook and of the central data object (Hatchet GraphFrame in this case) more explicit. Both the notebook and data states are complex and varied and screen space is limited, so further research is needed into making them more apparent to users.

\inlinehdr{The visualization and scripting paradigm suggests alternative designs for multiple coordinated views.}\ In our studies, we chose a straightforward, easy-to-understand use case of retrieving data from a visualization: recovering direct manipulation changes made to a dataset. We see this as an example of how to design coordinated view systems in a way that is consistent with how notebooks are already designed. Rather than overloading a single canvas with contextually relevant linked views, a visualization developer can provide these linked views for use when needed and to be placed in locations most relevant to the scripting side of the analysis. 

\section{Conclusion}
\label{sec:conclusion}
We conducted a design study investigating design concerns for visualization+scripting workflows. We proposed a space of tasks where their specificity, frequency, and suitability suggest \textit{how} they should be supported. Applying our methodology to
the context of performance analysis of calling context trees, we demonstrate how how these concerns are applied to several common abstract tree tasks. Through two participant studies, we observed our design anticipated participant needs and provided a viable path to expand the analysis beyond the visualization design as tasks evolved. We further observed a variety of strategies and queries used by our participants and their desire for a fluid, scripting-enhanced visual exploration workflow. However, our work also reveals challenges faced by participants regarding how they perceive and convey transitions between both data object state as well as notebook state. Further design work is needed to improve these transitions and state-tracking in notebook and similar contexts.


\section*{Acknowledgment}
This work was performed under the auspices of the U.S. Department of Energy by Lawrence Livermore National Laboratory under contract DE-AC52-07NA27344 as well as the United States Department of Defense through DTIC Contract FA8075-14-D-002-007, the National Science Foundation under NSF IIS-1844573 and IIS-2324465, and the Department of Energy under DE-SC0022044 and DE-SC0024635. LLNL-JRNL-859074.

\ifCLASSOPTIONcaptionsoff
  \newpage
\fi



%
\bibliographystyle{IEEEtran}

\bibliography{main}

%


\vspace{-40pt}
\begin{IEEEbiography}[{\includegraphics[width=1in,height=1.25in,clip,keepaspectratio]{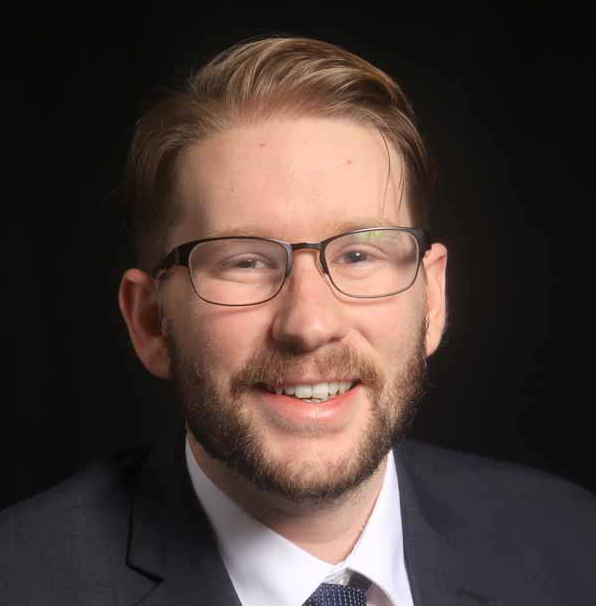}}]{Connor Scully-Allison} is a PhD student at the University of Utah. His research focuses on understanding the data and workflows used by data scientists for exploratory data analysis from a human-centered perspective.
\end{IEEEbiography}

\vspace{-40pt}
\begin{IEEEbiography}[{\includegraphics[width=1in,height=1.25in,clip,keepaspectratio]{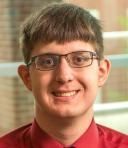}}]{Ian Lumsden} is a PhD student at the University of Tennessee, Knoxville advised by Dr. Michela Taufer. His research interests are developing novel techniques for HPC performance data analysis and developing tools to enable and enhance scientific computing workflows.
\end{IEEEbiography}

\vspace{-40pt}
\begin{IEEEbiography}[{\includegraphics[width=1in,height=1.25in,clip,keepaspectratio]{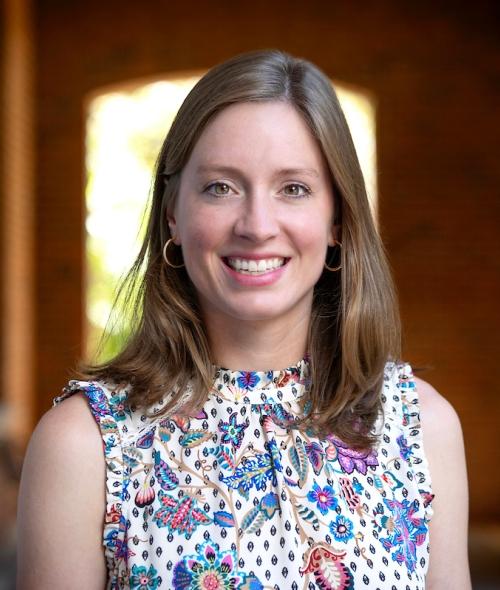}}]{Katy Williams} is an assistant professor of mathematics and computer science at Davidson College, who specializes in data visualization. Her research focuses on how people think about datasets and how best to translate those insights into visualization practices. 
\end{IEEEbiography}

\vspace{-40pt}
\begin{IEEEbiography}[{\includegraphics[width=1in,height=1.25in,clip,keepaspectratio]{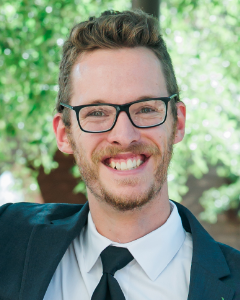}}]{Jesse Bartels} is a software developer that enjoys dynamic program analysis, data flow analysis, and efficient program execution tracing. He currently works at the Rincon Research corporation, learning new ways to make DSP run fast and interfacing next gen SDRs with general purpose CPUs.
\end{IEEEbiography}

\vspace{-40pt}
\begin{IEEEbiography}[{\includegraphics[width=1in,height=1.25in,clip,keepaspectratio]{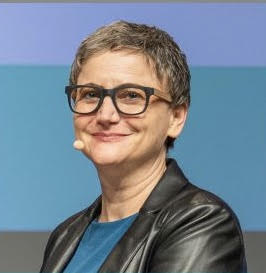}}]{Michela Taufer} holds the Jack Dongarra Professorship in High-Performance Computing within the Department of Electrical Engineering and Computer Science at the University of Tennessee, Knoxville. Dr. Taufer received her Ph.D. in computer science from the Swiss Federal Institute of Technology (ETH) in 2002. Her interdisciplinary research is at the intersection of computational sciences, high-permanence computing, and data analytics.
\end{IEEEbiography}

\vspace{-40pt}
\begin{IEEEbiography}[{\includegraphics[width=1in,height=1.25in,clip,keepaspectratio]{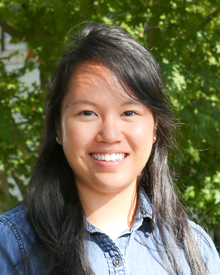}}]{Stephanie Brink} received her PhD degree in computer science from the University of Oregon, in 2019. She is a computer scientist with the Center for Applied Scientific Computing (CASC), Lawrence Livermore National Laboratory. Her research interests include involves developing scalable performance tools for controlling lowlevel hardware knobs as well as performance monitoring and analysis. Broadly, she is interested in power-constrained supercomputing, HPC system software, and performance tools.
\end{IEEEbiography}

\vspace{-40pt}
\begin{IEEEbiography}[{\includegraphics[width=1in,height=1.25in,clip,keepaspectratio]{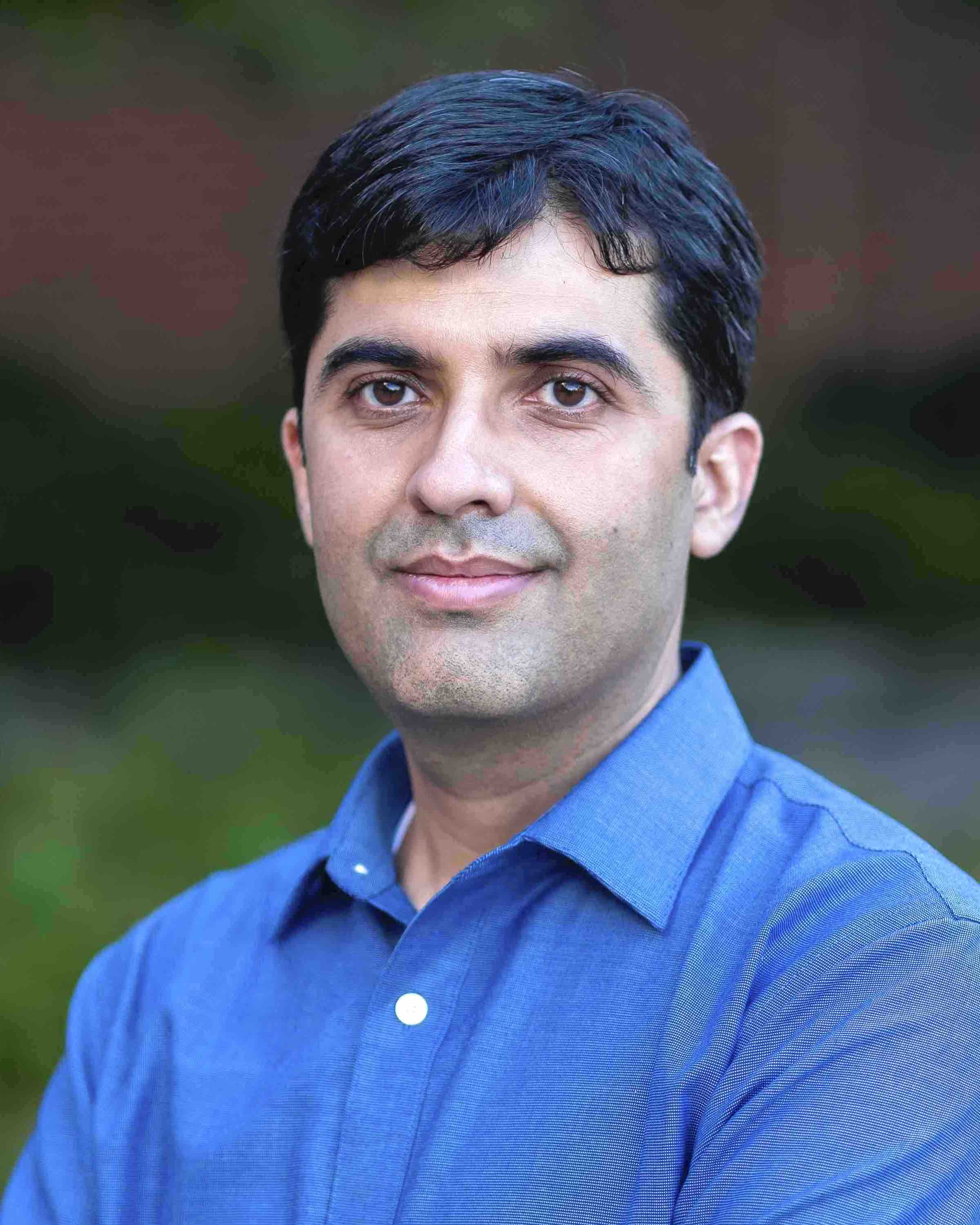}}]{Abhinav Bhatele} is an associate professor in the department of computer science, and director of the Parallel Software and Systems Group at the University of Maryland, College Park. His research interests are broadly in systems and AI, with a focus on parallel computing and distributed AI. He has published research in parallel programming models and runtimes, network design and simulation, applications of machine learning to parallel systems, parallel deep learning, and on analyzing/visualizing, modeling and optimizing the performance of parallel software and systems.\end{IEEEbiography}

\vspace{-35pt}
\begin{IEEEbiography}[{\includegraphics[width=1in,height=1.25in,clip,keepaspectratio]{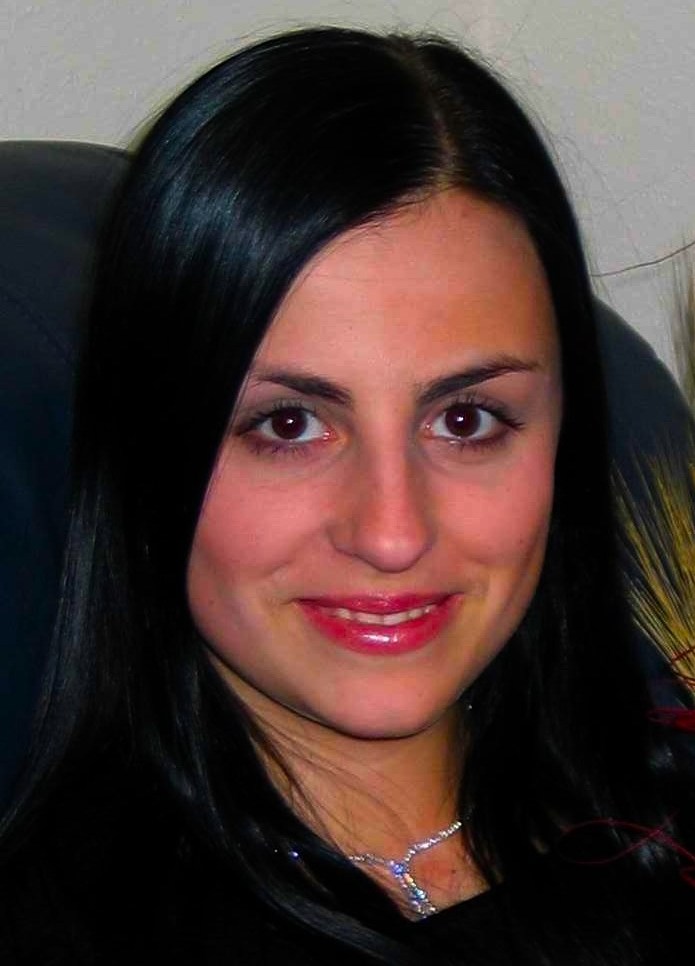}}]{Olga Pearce} is a computer scientist at Lawrence Livermore National Laboratory, where she serves as the benchmarking lead for LLNL’s Advanced Technology Systems, and the project PI for performance analysis.  Dr. Pearce holds a joint appointment with Computer Science and Engineering Department at Texas A\&M University.
\end{IEEEbiography}

\vspace{-35pt}
\begin{IEEEbiography}[{\includegraphics[width=1in,height=1.25in,clip,keepaspectratio]{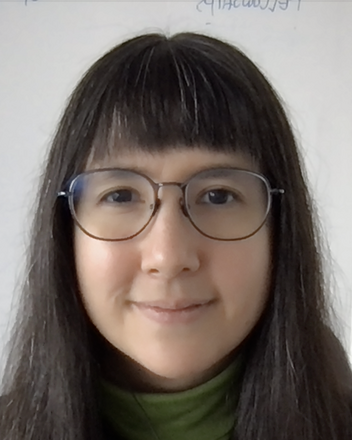}}]{Katherine E. Isaacs} is an associate professor in the Scientific Computing and Imaging (SCI) Institute and Kahlert School of Computing at the University of Utah. Her research focuses on data visualization challenges in complex exploratory analysis scenarios such as those of active research teams.
\end{IEEEbiography}





\end{document}